\documentclass[10pt,conference]{IEEEtran}
\IEEEoverridecommandlockouts
\usepackage{amssymb}
\usepackage{graphicx}
\usepackage{epstopdf}
\usepackage{multirow}
\usepackage{amsmath}
\usepackage{color,soul}
\usepackage{array}
\usepackage{url}
\usepackage[breaklinks]{hyperref}
\usepackage{breakurl}
\usepackage[export]{adjustbox}
\usepackage[T1]{fontenc}
\begin{document}
	\title{On the Formalization of Importance Measures using HOL Theorem Proving\\
	\thanks{The final publication is available at http://ieeexplore.ieee.org}}

\author{\IEEEauthorblockN{Waqar Ahmad$^{1}$,
Shahid Ali Murtza$^{2}$,
Osman Hasan$^{2}$, and
Sofi\`ene Tahar$^{1}$}
\IEEEauthorblockA{$^{1}$Electrical and Computer Engineering, \\
Concordia University, Montreal, QC, Canada \\
Email: \{waqar,tahar\}@ece.concordia.ca}
\IEEEauthorblockA{$^{2}$School of Electrical Engineering and Computer Science,\\
National University of Sciences and Technology, Islamabad, Pakistan\\
Email: \{smurtza.msee15seecs,osman.hasan\}@seecs.nust.edu.pk}
}	

\maketitle
	
\begin{abstract}
	Importance measures provide a systematic approach to scrutinize critical system components, which are extremely beneficial in making important decisions, such as prioritizing reliability improvement activities, identifying weak-links and effective usage of given resources. The importance measures are then in turn used to obtain a criticality value for each system component and to rank the components in descending manner. Simulations tools are generally used to perform importance measure based analysis, but they require expensive computations and thus they are not suitable for large systems. A more scalable approach is to utilize the importance measures to obtain all the necessary conditions by proving a generic relationship describing the relative importance between any pair of components in a system. In this paper, we propose to use higher-order-logic~(HOL) theorem proving to verify such relationships and thus making sure that all the essential conditions are accompanied by the proven property. In particular, we formalize the commonly used importance measures, such as Birnbaum and Fussell-Vesely, and conduct a formal importance measure analysis of a railway signaling system at a Moroccan level crossing as an application for illustration purpose.  
\end{abstract}


\IEEEpeerreviewmaketitle

\section{Introduction}
\label{sec:intro}
Importance measures~\cite{kuo2012importance} provide an effective way to evaluate the relative criticality of components in a system. Particularly, they are employed to identify a subset of components that are more important to a system so that given resources can be effectively utilized. The underline concept is to focus on the most problematic areas in a system aiming to achieve the most significant gains. A study at Microsoft Corp. has revealed that about 20\% of the entire pool of detected bugs lead to about 80\% of the errors and crashes in Microsoft Windows and Office software \cite{rooney2002microsoft}. In reliability engineering, determining the importance of components significantly helps to solve several reliability problems, such as component assignment, redundancy allocation, system upgrading, and fault diagnosis and maintenance. 

In 1968, Birnbaum was the first to propose the concept of \textit{importance measure} for binary systems of two states, either functioning or failed~\cite{birnbaum1968importance}. This led to the development of more sophisticated importance measures, such as Fussell-Vesely~\cite{kuo2012importance}, to analyze more complicated systems, like nuclear power plants. These importance measures are primarily defined for coherent systems~\cite{kuo2012importance}, which are systems satisfying the following conditions: (1) their structure function or system failure model exhibits non-decreasing behavior, i.e., the probability of the given failure model increases with the increase in the number of failures; and (2) each of their components is relevant, i.e., every component is actively contributing to the system failure.

A typical method in importance measure analysis involves calculating a criticality value for each component in a system and then tabulating the obtained data in descending manner~\cite{espiritu2007component}. In other words, a component with higher value is regarded as highly critical and placed above in the ranking than a component with a lower value. Simulation based reliability analysis tools, such as ReliaSoft~\cite{Reliasoft_RI}, determine the component's importance by computing the percentage of times that a system failure event is caused by a failure of a particular component over the simulation time 0 to \textit{t}. However, for analyzing the relative importance between all pairs of components, these methods have very high computational requirements especially when dealing with systems with many components. 

The scalability limitations of simulation based importance measure analysis can be resolved by using mathematically verified reduction methods in this context. For instance, Boland et al.~\cite{boland1989optimal} developed a relationship stating that the component \textit{i} is structurally more critical than the component \textit{j} if its structure function is larger when \textit{i} is down and \textit{j} is up as compared to the opposite case. This work is further extended by Meng~\cite{meng2004comparing} to obtain the necessary conditions, based on Birnbaum and Fussell-Vesely importance measures, that are essential for proving the analytical relationships describing the relative importance of any pair of system components. These analytical relationships can be extremely helpful in practical scenarios since calculating the exact values of a component importance measures can be tedious for large and complex systems. However, these analytical relationships have been manually verified using paper-and-pencil based proof methods and thus there is no guaranty that all necessary conditions are explicitly identified. This is a grave concern considering the safety-critical nature of some importance measure analysis. Thus, there is a dire need of developing more rigorous analysis of these foundational relationships to guarantee their correctness and their appropriate usage.

In this paper, we propose to utilize higher-order-logic (HOL) theorem proving to assure the formal guarantees about the relationships, obtained by Boland and Meng, governing the relative importance of any pair of system components. The HOL theorem prover is a system of deduction with precise semantics and provides a sound reasoning support for verifying the given properties, stated as a theorem, rigorously~\cite{harrison_09}. We first formalize the properties of coherent systems by describing their structure function as a fault tree (FT)~\cite{international2006iec} model, which is a graphical model for analyzing the conditions and factors causing the system failure.
Secondly, we formalize commonly used importance measures, such as Birnbaum, Fussell-Vesely, Reduction Worth and Achievement Worth~\cite{kuo2012importance}. We then use the formalization of Birnbaum importance measure to formally verify the relative importance properties of any pair of system components as described by Boland and Meng using HOL theorem proving. For illustration purposes, we conduct the formal importance measure analysis of a railway signaling system at a Moroccan level crossing (LC)~\cite{boudnnaya2014dependability} consisting of several critical components, such as lights, programmable logic controllers, alarms and also human factor, using the HOL theorem prover~\cite{hol_tp}.

The rest of the paper is organized as follows: An overview of the related work is presented in Section~\ref{sec:rel_work}. In Section~\ref{sec:prelim}, we provide a brief summary of the HOL theorem prover and the fundamentals of the HOL probability theory. A brief introduction to the recent formalization of FT analysis is also described to facilitate the understanding of the paper. Section~\ref{sec:Imp_measure} presents the HOL formalization of the concept of importance measure and its related properties.  Section~\ref{sec:imp_measure_LC} applies our proposed approach by describing the formal importance measure analysis of the signaling system at a Moroccan level crossing. Finally, Section~\ref{sec:concl} concludes the paper.

\section{Related Work}
\label{sec:rel_work}

Importance measure is a useful concept in reliability engineering and has been analyzed analytically~\cite{kuo2012importance} as well as using simulation tools~\cite{Reliasoft_RI}. The latter approach is practically adopted by industrial engineers due to their powerful features. These tools follow the typical approach of ranking the system components according to their criticality value. However, this approach requires high computations to obtain the criticality value for all system components and then perform the successive analysis, which may not be possible for large and complex systems. An alternate approach is to verify a relative measure relationship for any pair of components and obtain the necessary conditions, as described by Meng~\cite{meng2004comparing}.


Recently, a formal dependability analysis framework~\cite{ahmad_phd_17}, based on Reliability Block Diagram~\cite{Trivedi_02,ahmed2016formalization} and FT modeling techniques, has been developed using HOL theorem proving. This framework has been successfully utilized to carry out the reliability analysis of a railway traction drive system~\cite{ahmad_railway_trac}, failure analysis of satellite solar arrays~\cite{CICM_15_WAhmed} and an air traffic management system~\cite{WAhmad_SETTA}.  In the current work, we formalize the notion of coherent system and the importance measure by representing the system structure function based on existing FT models. To the best of our knowledge, this is the first formal work describing the formalization of the importance measures using HOL theorem proving.

\section{Preliminaries}
\label{sec:prelim}
In this section, we first give a brief introduction to the HOL theorem prover, formalization of probability theory and an approach for the formal FT analysis to facilitate the understanding of the rest of the paper.

\subsection{HOL Theorem Prover}

HOL~\cite{gordon1993introduction} is an interactive theorem prover, developed at the
University of Cambridge, UK, for conducting proofs in higher-order logic.
It utilizes the simple type theory of Church \cite{church_40} along
with Hindley-Milner polymorphism \cite{milner_77} to implement
higher-order logic. HOL has been successfully used as a verification
framework for both software and hardware as well as a platform for
the formalization of pure mathematics.

The HOL core consists of only 4 basic
axioms and 8 primitive inference rules, which are implemented as ML
functions. The ML's type system ensures that only valid theorems can be constructed. Soundness is assured as every new theorem must be verified by applying these basic axioms and primitive inference rules or any other previously verified theorems/inference rules.

In this paper, we utilize the HOL theories (libraries) of
Booleans, lists, sets, positive integers, \emph{real} numbers,
measure and probability~\cite{mhamdi_11}. In fact, one of the primary
motivations of selecting the HOL theorem prover for our work was to
benefit from these built-in mathematical theories.
Table \ref{hol_basics} provides the mathematical interpretations of
some frequently used HOL symbols and functions, which are inherited
from existing HOL theories.

\begin{table}[!htb]
	\begin{center}
		\caption{HOL Symbols and Functions}
		\begin{tabular}{|c|c|c|} \hline
			{\bfseries HOL Symbol} & {\bfseries Standard Symbol} & {\bfseries Meaning}  \\
			\hline \hline
			$\mathtt{\wedge}$& $and$  & Logical $and$   \\ \hline
			$\mathtt{\vee}$ & $or$  &  Logical $or$  \\ \hline
			$\mathtt{\neg}$ & $not$  &  Logical $negation$  \\ \hline
			$\mathtt{::}$  &  $cons$ & Adds a new element to a list  \\ \hline
			$\mathtt{++}$  &  $append$ & Joins two lists together   \\ \hline
			$\mathtt{HD\ L}$  &  $head$ & Head element of list $L$ \\ \hline
			$\mathtt{TL\ L}$  &  $tail$ & Tail of list $L$\\ \hline
			$\mathtt{EL\ n\ L}$  &  $element$ & $n^{th}$ element of list L \\ \hline
			$\mathtt{MEM\ a\ L}$  &  $member$ & True if $a$ is a member of list $L$\\ \hline
			$\mathtt{\lambda x.t}$& $\lambda x.t$  &  Function that maps $x$ to $t(x)$  \\ \hline
			$\mathtt{SUC\ n}$& $n + 1$  &  Successor of a $num$  \\ \hline
			\hline
		\end{tabular}
		\label{hol_basics}
	\end{center}
\end{table}


\subsection{Probability Theory}
\label{subsec:prob_FT_hol}
Mathematically, a measure space is defined as a triple ($\Omega,\Sigma, \mu$), where
$\Omega$ is a set, called the sample space, $\Sigma$ represents a $\sigma$-algebra of subsets of
$\Omega$, where the subsets are usually referred to as measurable sets, and $\mu$ is a measure with domain
$\Sigma$. A probability space is a measure space ($\Omega,\Sigma, Pr$), such that the measure,
referred to as the probability and denoted by $Pr$, of the sample space is 1. In the HOL formalization of probability theory \cite{mhamdi_11}, given a probability space $p$, the functions \texttt{space}, \texttt{subsets} and \texttt{prob} return the corresponding $\Omega$, $\Sigma$ and $Pr$, respectively. This formalization also includes the formal verification of the commonly used probability laws, which play a pivotal role in formal reasoning about dependability properties.

A random variable is a measurable function between a probability space and a measurable space. The measurable functions belong to a special class of functions, which preserve the property that the inverse image of each measurable set  is also measurable. A measurable space refers to a pair ($S,\mathcal{A}$), where $S$ denotes a set and $\mathcal{A}$ represents a nonempty collection of sub-sets of $S$. Now, if $S$ is a set with finite elements, then the corresponding random variable is termed as a discrete otherwise it is called continuous.

The cumulative distribution function (CDF) is defined as the probability of an event where a random variable $X$ has a value less than or equal to some value $t$, i.e., $Pr(X \le t)$. This definition characterizes the distribution of both discrete and continuous random variables and has been formalized \cite{WAhmad_CICM14}:

\begin{flushleft}
	\label{CDF_def}
	\vspace{1pt} \texttt{$\vdash$ $\forall$  p X t. CDF p X t =\\ distribution p X \{y | y $\leq$ Normal t\}
	}
\end{flushleft}

\noindent The function \texttt{Normal} takes  a $real$ number as its input and converts it to its corresponding value in the $extended$-$real$ data-type, i.e, it is the $real$ data-type with the inclusion of  positive and negative infinity. The function \texttt{distribution} takes three parameters:  a probability space $p$, a random variable $X: (\alpha \rightarrow extreal)$ and a set of $extended$-$real$ numbers and returns  the probability of the given random variable $X$ acquiring  all the values of the given set in probability space.

The unreliability or the probability of failure $F(t)$ is defined as the probability that a system or component will fail by the time $t$. It can be described in terms of CDF, known as the failure distribution function, if a random variable $X$ represents the time-to-failure of the component. This time-to-failure random variable $X$ usually exhibits the exponential or Weibull distribution.

The notion of mutual independence of $n$ random variables is a major requirement for reasoning about the failure analysis of the given systems. According to this notion, a list of $n$ events are mutual independent if and only if for each set of $k$ events, such that $\mathit(1 \le k \le n)$, we have:
\begin{equation}\label{eq1:mutual_indep}
Pr(\bigcap_{i=1}^{k}A_i) = \prod_{i=1}^{k} Pr(A_i)
\end{equation}

The mutual independence concept has been formalized in the HOL theorem prover and more details can be found in~\cite{WAhmad_CICM14}.

\subsection{Fault Trees}
\label{subsec:formal_FTs}

Fault Tree (FT) analysis is a widely used technique to determine the dependability of real-world systems, like railways signaling, automotive or avionics.  It mainly provides a graphical model for analyzing the conditions and factors causing an undesired top event, i.e., a critical event, which can cause the complete system failure upon its occurrence. The preceding nodes of the FT are represented by gates, like OR, AND and XOR, which are used to link two or more cause events of a fault in a prescribed manner. 


The FT gates are formally modeled by using a  new recursive datatype $gate$ in HOL as follows~\cite{WAhmad_SETTA}:

\vspace{20mm}
\begin{flushleft}
	\texttt{\small   Hol\_datatype `gate = AND of gate list |\\
		\quad \qquad \qquad \quad \qquad \qquad \quad \quad \ OR of gate list |\\
		 \quad \qquad \qquad \quad \qquad \qquad \quad \quad \ NOT of gate |\\
		\quad \qquad \qquad \quad \qquad \qquad \quad \quad \  atomic of 'a  event`}
\end{flushleft}

\noindent The type constructors \texttt{AND} and \texttt{OR} recursively function on \textit{gate}-typed lists and the type constructor \texttt{NOT} operates on \textit{gate}-type variable. The type constructor \texttt{atomic} is basically a typecasting operator between \textit{event} and \textit{gate}-typed variables. 

A semantic function  is then defined over \textit{gate} datatype that can yield the corresponding event from the given FT gate as follows:

\begin{flushleft}
	\label{rbd_struct}
	\small
	\vspace{1pt} \texttt{$\vdash$
		($\forall$ p. FTree p (AND []) = p\_space p) $\wedge$\\
		\ \ ($\forall$ xs x p.\\
		\ \ \ \ FTree p (AND (x::xs)) =\\ 
		\ \ \ \ FTree p x $\cap$ FTree p (AND xs)) $\wedge$\\
		\ \ ($\forall$ p. FTree p (OR []) = \{\}) $\wedge$\\
		\ \ ($\forall$ xs x p.\\
		\ \ \ \ FTree p (OR (x::xs)) =\\ 
		\ \ \ \ FTree p x $\cup$ FTree p (OR xs))  $\wedge$ \\
		\ \ ($\forall$ p a.\\
		\ \ \ \ FTree p (NOT a) =\\
		 \ \ \ \ p\_space p DIFF FTree p a) $\wedge$\\
		\ \ ($\forall$ p a. FTree p (atomic a) = a)
	}
\end{flushleft}

\noindent The function \texttt{FTree} takes a list of type \emph{gate},  identified by the type constructor \texttt{AND}, and returns a complete probability space \texttt{p\_space p} if a given list is empty and otherwise returns the intersection of events in a given list. Similarly, to model the behavior of the OR FT gate, the function \texttt{FTree} returns the union of all the events after applying the function \texttt{FTree} on each element of a given list or an empty set if a given list is empty. The function \texttt{FTree} takes the type constructor \texttt{NOT} and returns the complement of a failure event obtained from the function \texttt{FTree}. The function \texttt{FTree} returns the failure event using the type constructor \texttt{atomic}.

If the occurrence of a failure event at the output is caused by the occurrence of all the input failure events, then this kind of behavior can be modeled by using the AND FT gate. Similarly, in the OR FT gate, the occurrence of an output failure event depends upon the occurrence of any one of its input failure event. The NOT FT gate can be used in conjunction with the AND and OR FT gates to formalize other FT gates. The formalization of these gates is based on~\cite{WAhmad_SETTA}, given in Table \ref{table:FT_gate_def}. The NAND FT gate, represented by the function \texttt{NAND\_FT\_gate} in Table \ref{table:FT_gate_def}, models the behavior of the occurrence of an output failure event when at least one of the failure events at its input does not occur. This type of gate is used in FTs when the non-occurrence of a failure event in conjunction with other failure events cause the top failure event to occur. This behavior can be expressed as the intersection of complementary and normal events, where the complementary events model the non-occurring failure events and the normal events model the occurring failure events. The output failure event occurs in the 2-input XOR FT gate if only one, and not both, of its input failure events occur.

\begin{table}[!htb]
	\centering
	\caption{HOL Formalization of Fault Tree Gates}
	\scalebox{0.75}{
		\begin{tabular}{|m{17mm} |m{91mm}|}
			\hline
			FT Gates & Formalization  \\
			\hline
			\hline
			\includegraphics[valign=c,scale=0.3,trim={0cm 0cm 0 0cm},clip]{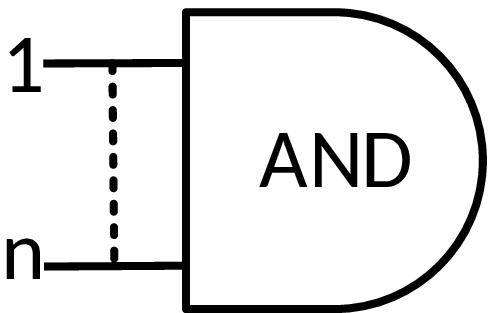} & 	
			$\!\begin{aligned}[c]
			&\small{\texttt{$\vdash$ $\forall$ p L1 L2.}}\\
			& \small{ \texttt{AND\_FT\_gate p L1 L2 = }} 
			\small{ \texttt{FTree p (AND (gate\_list L))}
			}\end{aligned}$ \\
			\hline
			\includegraphics[valign=c,scale=0.3,trim={0cm 0cm 0 0cm},clip]{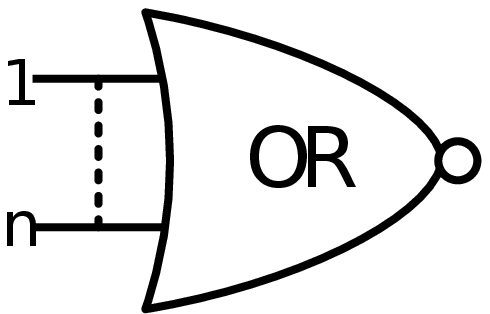}&
			$\!\begin{aligned}[c]
			&\small{\texttt{$\vdash$ $\forall$ p L.}}\\
			& \small{ \texttt{OR\_FT\_gate p L = }}
			\small{ \texttt{FTree p (OR (gate\_list L))}
			}\end{aligned}$ \\
			\hline		
			\includegraphics[valign=c,scale=0.3,trim={0cm 0cm 0 0cm},clip]{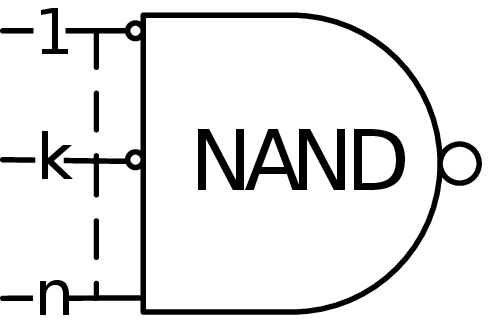} & 	
			$\!\begin{aligned}[c]
			&\small{\texttt{$\vdash$ $\forall$ p L1 L2.}}\\
			& \small{ \texttt{NAND\_FT\_gate p L1 L2 = }}\\
			&  \small{ \texttt{FTree p (AND (gate\_list}}
			\small{ \texttt{(compl\_list p L1 ++ L2)))}
			}\end{aligned}$ \\
			\hline
			\includegraphics[valign=c,scale=0.3,trim={0cm 0cm 0 0cm},clip]{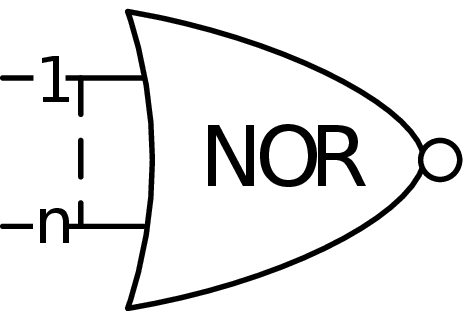}&
			$\!\begin{aligned}[c]
			&\small{\texttt{$\vdash$ $\forall$ p L.}}\\
			& \small{ \texttt{NOR\_FT\_gate p L =}}\\
			& \small{ \texttt{FTree p (NOT (OR (gate\_list L)))}
			}\end{aligned}$ \\
			\hline		
			\includegraphics[valign=c,scale=0.3,trim={0cm 0cm 0 0cm},clip]{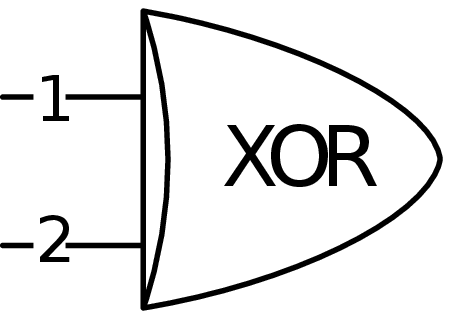} &
			$\!\begin{aligned}[c]
			& \small{\texttt{$\vdash$ $\forall$ p A B.}}\\
			&\small{ \texttt{XOR\_FT\_gate p A B =}}\\
			& \small{\texttt{FTree p }}
			\small{ \texttt{(OR [AND [NOT A; B]; AND [A; NOT B]])
			}}\end{aligned}$ \\
			
			\hline
	\end{tabular}}\label{table:FT_gate_def}
\end{table}

The verification of the corresponding failure probability expressions, of the above-mentioned FT gates, is presented in Table \ref{FT_exp_table}. These expressions are verified under the following assumptions: (i) \texttt{prob\_space p} ensures that $p$ is a valid probability space; (ii) \texttt{2  $\le$ LENGTH L} makes sure that the given list $L$ must have at least two elements; (iii) \texttt{in\_events p L} ensures that all the corresponding events in the given list $L$ are drawn from the events space $p$; and (iv) \texttt{mutual\_indep p L} guarantees that events in the given list~$L$ are mutually independent~\cite{WAhmad_CICM14}.		

\begin{table}[!htb]
	\centering
	\caption{Probability of Failures of Fault Tree Gates}
	\scalebox{0.8}{
		\begin{tabular}{|m{5cm}|m{5.2cm}|}
			\hline
			Mathematical Expressions & Theorem's Conclusion \\
			\hline
			\hline
			$\!\begin{aligned}[c]
			F_{AND}(t) & = Pr (\bigcap_{i=2}^{N}A_{i}(t))\\
			& = \prod_{i=2}^{N}F_{i}(t)
			\end{aligned}$  &
			$\!\begin{aligned}[c]
			& \small{\texttt{$\vdash$ $\forall$ p L1 L2.}}\\ & \small{\texttt{ (prob p (AND\_FT\_gate L) =}}\\
			 &\small{\texttt{\ \ list\_prod (list\_prob p L))}}
			\end{aligned}$ \\
			\hline
			$\!\begin{aligned}[c]
			F_{OR}(t) & = Pr (\bigcup_{i=2}^{N}A_{i}(t))\\
			& = 1 - \prod_{i=2}^{N}(1 - F_{i}(t))
			\end{aligned}$  &
			$\!\begin{aligned}[c]
			& \small{\texttt{$\vdash$ $\forall$ p L1 L2.}}\\ 
			& \small{\texttt{(prob p (OR\_FT\_gate p L) =}}\\
			  &\small{\texttt{1 - list\_prod(one\_minus\_list}}\\
			  & \small{\texttt{ \ \  (list\_prob p L)))}}
			\end{aligned}$ \\
			\hline
			$\!\begin{aligned}[c]
			F_{NAND}(t)
			& =  Pr (\bigcap_{i=2}^{k}\overline A_{i}(t) \cap \bigcap_{j=k}^{N}A_{i}(t)) \\
			&= \prod_{i=2}^{k}(1 - F_{i}(t)) *\prod_{j=k}^{N}(F_{j}(t))
			\end{aligned}$  &
			$\!\begin{aligned}[c]
			& \small{\texttt{$\vdash$ $\forall$ p L1 L2.}}\\ 
			& \small{\texttt{(prob p}}\\
			& \small{\texttt{ \ \ (NAND\_FT\_gate p L1 L2)  =}}\\
			& \ \small{\texttt{list\_prod (list\_prob p}}\\
			& \small{\texttt{ \ \ (compl\_list p L1)) *}}\\
			& \small{\texttt{ list\_prod (list\_prob p L2))}}
			\end{aligned}$ \\
			\hline
			$\!\begin{aligned}[c]
			F_{NOR}(t)
			& = 1 - F_{OR}(t)  = \prod_{i=2}^{N}(1 - F_{i}(t))
			\end{aligned}$  &
			$\!\begin{aligned}[c]
			& \small{\texttt{$\vdash$ $\forall$ p L.}}\\
			& \small{\texttt{ (prob p (NOR\_FT\_gate p L)  =}} \\
			& \ \  \small{\texttt{list\_prod (one\_minus\_list}}\\
			& \small{\texttt{ \ \ (list\_prob p L)))}}
			\end{aligned}$ \\
			\hline
			$\!\begin{aligned}[c]
			F_{XOR}(t)
			&= Pr(\overline{A}(t)B(t) \cup A(t)\overline{B}(t)) \\
			&= (1- F_{A}(t))F_{B}(t) +\\
			& \ \ \ \    F_{A}(t)(1- F_{B}(t))
			\end{aligned}$  &
			$\!\begin{aligned}[c]
			& \texttt{$\vdash$ $\forall$ p A B.}\\
		 & \texttt{prob\_space p $\wedge$ A $\in$ events p $\wedge$}\\
		 & \texttt{B $\in$ events p $\Rightarrow$}\\
			& \texttt{(prob p (XOR\_FT\_gate p A B) = }\\
			& \texttt{(1 - prob p A) * prob p B +}\\
			& \texttt{\ \ prob p A * (1 - prob p B))}
			\end{aligned}$ \\

			\hline
	\end{tabular}}\label{FT_exp_table} \vspace*{-0.2cm}
\end{table}

\subsection{PIE Principle }
\label{subsec:PIE}							
In FT analysis, the first step is to identify all the basic  failure events that can cause the occurrence of the system top failure event. These failure events are then combined to model the overall fault behavior of a given system by using the fault gates. These combinations of basic failure events, called cut sets, are then reduced to  minimal cut sets (MCS) by using set-theory rules, such as idempotent, associative and commutative. Then, the Probabilistic Inclusion Exclusion~(PIE) principle is used to evaluate the overall failure probability of a given system based on the MCS events. According to the PIE principle, if $A_{i}$ represents the $i^{th}$ basic failure event or a combination of failure events, then the overall failure probability of a given system can be expressed as follows:

\begin{equation}\label{PIE}
\mathbb{P} (\bigcup_{i=1}^n A_i)  = \sum_{t \neq \{\}, t\subseteq\{1,2,\ldots,n\}}(-1)^{|t|+1} \mathbb{P} (\bigcap_{j\in t} A_j)
\end{equation}

\noindent The above equation has been formally verified in HOL and details can be found in~\cite{WAhmad_SETTA}.
%

\section{Importance Measures} 
\label{sec:Imp_measure}
The concept of importance measure is proposed by Birnbaum mainly for components of coherent systems. This section describes the essential properties of a coherent system that is then followed by the commonly used importance measures and their respective formalizations in HOL.

\subsection{Coherent System}
\label{subsec:coherent_sys}
Let, $\phi(\bar{x})$ be the structure function of a system functioning on the \textit{n}-components state vector $\bar{x}=(x_1, x_2, \ldots, x_i, \ldots, x_n)$, where $x_i$ is the state of the $i^{th}$ component. According to Birnbaum~\cite{birnbaum1968importance}, a system of binary state, where both the system and its components can either be in state of failure or success, is said to be coherent if its structure function, $\phi(\bar{x})$, satisfies the following conditions:

\begin{enumerate}
	\item[(1)] $\phi(\bar{0})= 0$ with $\bar{0}=(0,0,\ldots,0)$\\
	\item[(2)] $\phi(\bar{1})= 1$ with $\bar{1}=(1,1,\ldots,1)$\\
	\item[(3)] $\phi(\bar{x}) \le \phi(\bar{y})$ if $\bar{x} \le \bar{y}$ with relationship $\bar{x} \le \bar{y}$ means $ x_i \le y_i, \forall i=1, 2, ... n$. 
\end{enumerate}

The first two conditions state that a system must go in state 0 (full working) or 1 (complete failure) if all of its components are in state 0  or 1, respectively. The third condition defines the monotonicity property of a system structure function ensuring that a component in working state must not contribute in causing a system failure and vice versa. The use of the NOT gate in a FT model (structure function) results in a non-coherent structure, which also means that components not  failing,  i.e., working,  can  contribute  to  a  system  failure  event and thus violating the condition (3). Therefore, the use of the NOT logic is often discouraged~\cite{andrews2003birnbaum}.

In order to formally verify that a given failure structure function (FT model) satisfies the Birnbaum coherent system conditions, we first formally define a structure function in HOL as follows: 

\begin{flushleft}
	\texttt{\bf{Definition 1: }}
	\label{def:coherent_struct}
	\vspace{1pt} \small{\texttt{$\vdash$ $\forall$ f L. $\phi$ f L = f L
	}}
\end{flushleft}

\noindent where $\phi$ is a HOL Unicode character that is used as a pretty-printing of the function \texttt{coherent\_struct} mapping an arbitrary real-valued function $f:(\alpha\rightarrow bool) \rightarrow real$ to a list of sets $L: (\alpha \rightarrow bool) list$. Using the above definition, Conditions (1-2) can be verified in HOL on a given fault tree (structure function) consisting of AND and OR FT gates as:
\vspace{5mm}
\begin{flushleft}
	\texttt{\bf{Theorem 1: }}
	\label{thm:coherent_struct_C1}
	\vspace{1pt} \small{\texttt{$\vdash$ $\forall$ p L.\\
			prob\_space p  $\wedge$ $\neg$NULL L $\wedge$ \\
			coherent\_state\_vec ($\lambda$a. a = \{\}) (FLAT L) $\Rightarrow$\\
			(prob p\\
			\ \ ($\phi$ ($\lambda$b.\\
			 \ \ \ \ \ \ FTree p ((OR of\\
			  \  \ \ \ \ \ \ \ ($\lambda$a. AND (gate\_list a))) b)) L) = 0)
	}}
\end{flushleft}

\begin{flushleft}
	\texttt{\bf{Theorem 2: }}
	\label{thm:coherent_struct_C2}
	\vspace{1pt} \small{\texttt{$\vdash$ $\forall$ p L.\\
			prob\_space p  $\wedge$ $\neg$NULL L $\wedge$\\
			coherent\_state\_vec\\
			\ \  ($\lambda$a. a = p\_space p) (FLAT L) $\Rightarrow$\\
			(prob p\\
			\ \ ($\phi$ ($\lambda$b.\\
			\ \ \ \ \ \  FTree p ((OR of\\
			\  \ \ \ \ \ \ \   ($\lambda$a. AND (gate\_list a))) b)) L) = 1)
	}}
\end{flushleft}

\noindent where, the HOL function \texttt{FLAT} is used to flatten the two-dimensional list, i.e., to transform a list of lists, into a single list. The assumptions in the above theorems are almost similar. The first two assumptions ensure that the variable~\textit{p} is a valid probability space and the given list of state vectors is not empty. In the last assumption, the function \texttt{coherent\_state\_vec} asserts that all the system components are either in fully working or in completely failure state, which are modeled using empty event (\{\}) and the complete probability space (\texttt{p\_space p}), respectively. The conclusions of the above theorems model the probabilistic sense of the conditions (1-2).

Now, we formally verify the Condition 3 for the given structure function in HOL as follows:
\begin{flushleft}
	\texttt{\bf{Theorem 3: }}
	\label{thm:coherent_struct_C3}
	\vspace{1pt} \small{\texttt{$\vdash$ $\forall$ p L.\\
			prob\_space p  $\wedge$ \\
			in\_events p (FLAT (XL\_vec L)) $\wedge$ \\
			in\_events p (FLAT (YL\_vec L))  $\wedge$ \\ mem\_subset\_vec L $\Rightarrow$\\
			prob p\\
			\ ($\phi$ ($\lambda$b.\\
			\ \ \ FTree p((OR of\\
			\ \ \  \ ($\lambda$a. AND (gate\_list a))) b)) (XL\_vec L)) $\le$\\
			prob p\\
			\ ($\phi$ ($\lambda$b.\\
			\ \ \ \ FTree p ((OR of\\
			\ \ \ \ \ \ ($\lambda$a. AND (gate\_list a))) b)) (YL\_vec L))
	}}
\end{flushleft}

\noindent where the function \texttt{in\_events} ensures that each element of a given list belongs to a valid event space \textit{p}. The functions \texttt{XL\_vec} and \texttt{YL\_vec} returns the first and second member of the two-dimensional pair list, respectively. The relationship between these two lists,  \texttt{XL\_vec} and \texttt{YL\_vec}, is described by the function \texttt{mem\_subset\_vec}, which ensures that each member of \texttt{XL\_vec} list is a subset of the corresponding member of \texttt{YL\_vec} list. The conclusion of the above theorem models Condition~3. The proof of Theorem 3 follows from the fact that if $A \subseteq B$, then their corresponding probabilities satisfy the monotonicity property, i.e., $\Pr(A) \le \Pr(B)$.



\subsection{Birnbaum Importance} 

For a coherent system of \textit{n}-components with independent failures, the Birnbaum importance ($I^{(i)}_{B}$) of component $i$ is defined as a probability that the $i^{th}$ component is critical to the system failure or functioning. 
Mathematically, it can be expressed as follows~\cite{birnbaum1968importance}:

\begin{equation}
\label{eq:rel_import0}
\dfrac{\partial h (\bar{x})}{\partial p_i} = I_{B}^{(i)} (\phi(\bar{x}))= \Pr \{\phi((1_i,\bar{x}))\} - \Pr \{\phi((0_i,\bar{x}))\}
\end{equation}
 
\noindent where $\phi(\bar{x})$ represents the structure function of a given coherent system, which is applied on components state vector $\bar{x}$ and returns the corresponding state of a system. 
The notations $\phi((1_i,\bar{x}))$ and $\phi((0_i,\bar{x}))$ represent the state of a system if the $i^{th}$ component is updated with the state values 1 (failure) and 0 (working), respectively. 

To formalize Equation \ref{eq:rel_import0} in HOL, we first formally define the notion of component state update in a given structure function as follows:

\begin{flushleft}
	\texttt{\bf{Definition 2: }}
	\label{def:coherent_struct_update}
	\vspace{1pt} \small{\texttt{$\vdash$ $\forall$ i f L.\\
			 $\phi$' e i f L = $\phi$ f (LUPDATE e i L)
	}}
\end{flushleft}

\noindent where the HOL function \texttt{LUPDATE} updates the given list \textit{L} with element \textit{e} at index \textit{i}. The above function updates the state of the component  \textit{i} in a state vector \textit{L} before passing it to the system structure function. Similarly, we can formally define a function to update the states of any two system components in HOL as follows:

\begin{flushleft}
	\texttt{\bf{Definition 3: }}
	\label{def:coherent_struct_update2}
	\vspace{1pt} \small{\texttt{$\vdash$ $\forall$ e e' i j f L.\\
			$\phi$'' e e' i j f L = \\
			$\phi$ f (LUPDATE e' j (LUPDATE e i L))
	}}
\end{flushleft}

Now, using Definition 2, we can formally model Equation~\ref{eq:rel_import0} in HOL as follows:

\begin{flushleft}
	\texttt{\bf{Definition 4: }}
	\vspace{1pt} \small{\texttt{$\vdash$ $\forall$ p i f L.\\
			$I_\beta$ p i f L =\\
			prob p ($\phi$' (p\_space p) i f L) -\\
			prob p ($\phi$' \{\} i f L)
	}}
\end{flushleft}

As described earlier in Section~\ref{sec:intro}, Meng~\cite{meng2004comparing} developed the analytical relationship describing the relative importance of any pair of system components and obtained the necessary conditions based on Boland and Birnbaum importance measures. We formally verify this relationship in HOL as follows:


\begin{flushleft}
	\texttt{\bf{Theorem 4: }}
	\label{thm:BImp_rel}
	\emph{Meng~\cite{meng2004comparing}: Suppose that \textit{i} $\stackrel{c}{=}$ \textit{j} and $\dfrac{\partial^2 h(\textbf{x})}{\partial p_i \partial p_j}$ $\ge$ 0 for all \textbf{x}. Then, $I_{\beta}$(j,\textbf{x}) $\le$  $I_{\beta}$(i,\textbf{x}) for all \textbf{x} satisfying $p_i$ $\le$ $p_j$. } \\
	\vspace{3mm}
	\vspace{1pt} \small{\texttt{$\vdash$ $\forall$ p L i j.\\ \
			[A1]: prob\_space p $\wedge$ in\_events p L $\wedge$ \\ \
			[A2]: $\neg$NULL L $\wedge$  i < j $\wedge$\\		\	
			[A3]: mutual\_indep p\\
			\ \ \ \ \ \ \ (\{\} :: \{\} ::p\_space p::p\_space p::L) $\wedge$\\ \
			[A4]: SUC (SUC j) < LENGTH L $\wedge$\\ \
			[A5]: $I''_\beta$ p i j\\
			\ \ \ \ \ ($\lambda$a. FTree p (AND (gate\_list a))) L $\ge$ 0 $\wedge$\\ \
			[A6]: prob p (EL i L) $\le$ prob p (EL j L) $\Rightarrow$\\
			($I_\beta$ p j ($\lambda$a. FTree p (AND (gate\_list a))) L $\le$\\
			\ $I_\beta$ p i ($\lambda$a. FTree p (AND (gate\_list a))) L)
	}}
\end{flushleft}

\noindent 
In the above statement, the symbol \textit{i} $\stackrel{c}{=}$ \textit{j} is described by Boland et al.~\cite{boland1989optimal} as 
components \textit{i} and \textit{j} are permutation equivalent if $\phi(1_i,0_j,\textbf{x}) = \phi(0_i,1_j,\textbf{x})$ for all~\textbf{x}. Using Definition 3, we formally verify this property in HOL as follows:

\begin{flushleft}
	\texttt{\bf{Lemma 1: }}
	\vspace{1pt} \small{\texttt{$\vdash$ $\forall$ p i j L. prob\_space p $\wedge$  i < j $\wedge$\\
			in\_events p L $\wedge$ SUC (SUC j) < LENGTH L $\wedge$\\
			mutual\_indep p (\{\} ::p\_space p::L) $\Rightarrow$\\
			(prob p ($\phi$'' (p\_space p) \{\} i j\\
			\ \ \ \ ($\lambda$a. FTree p (AND (gate\_list a))) L) =\\
			\ prob p ($\phi$'' \{\} (p\_space p) i j\\
			\ \ \ \ ($\lambda$a. FTree p (AND (gate\_list a))) L))
	}}
\end{flushleft}

\noindent Similarly, the notation  $\dfrac{\partial^2 h(x)}{\partial p_i \partial p_j}$ is a partial differentiation w.r.t probability of components \textit{i} and \textit{j} that can be  represented mathematically as:

\begin{equation}
	\begin{aligned}
	&\dfrac{\partial^2 h(x)}{\partial p_i \partial p_j} = \Pr(\phi^{''}(1_i,1_j,\textbf{x})) - \Pr(\phi^{'}(1_i,0_j,\textbf{x})) -\\
	& \ \ \ \ \ \ \quad \quad \Pr(\phi^{''}(0_i,1_j,\textbf{x})) + \Pr(\phi^{''}(0_i,0_j,\textbf{x}))
	\end{aligned}
\end{equation} 

\noindent The above equation is formalized using Definition 3 and it is represented by the function \texttt{$I''_\beta$} in the assumption (A5) of Theorem 4.

The assumptions of Theorem 4 are similar to the ones used in Theorems~1-3. The inclusion of \{\} and \texttt{p\_space p} in assumption (A3) reflects the change caused by flipping the state of the $i$ and $j$ components and also makes sure that they are mutually independence. The assumption (A4) ensures that the index $j$ starts after two increments since we require at least two components in a list.  Although a brief proof sketch of Theorem~4 is described by Meng ~\cite{meng2004comparing}, the sound environment of the HOL theorem prover provides additional formal guarantees in the verification of Theorem 4 accompanying all the necessary conditions. The formal proof of Theorem 4 utilizes several essential lemmas, which can be found in~\cite{B_Imp_19}.


\subsection{Other Common Types of Importance Measures}
Another well-known importance measure is Fussell-Vesely~\cite{kuo2012importance}, which describes the importance of component~\textit{i} as a probability that the failure of component~\textit{i} contributes to a system failure given that system fails. It can be expressed mathematically as follows:

 \begin{equation}
 \label{eq:FV_imp}
 	I_{FV} = \frac{\Pr (\phi(\textbf{x})) - \Pr(\phi'(1_i,\textbf{x}))}{\Pr(\phi(\textbf{x}))}
 \end{equation}
 
 We can formally define the above function by using Definitions 1-2 in HOL as follows:
 
 \begin{flushleft}
 	\texttt{\bf{Definition 5: }}
 	\vspace{1pt} \small{\texttt{$\vdash$ $\forall$ f i L.\\
 			I\_FV p i f L = \\	\vspace{2mm}		$\dfrac{\texttt{prob p ($\phi$ f L) - prob p ($\phi'$ (p\_space p) i f L)}} {\texttt{prob p ($\phi$ f L)}}$ 			
 	}}
 \end{flushleft}

 Similarly, the criticality importance measures Reducation Worth ($I_{RW}$) and Achievement Worth ($I_{AW}$) describe a probability when component \textit{i} is always functioning and failed, respectively. They can be expressed as follows:
 
 \begin{equation}
 \label{eq:AW_RW_imp}
 \begin{aligned}
 I_{RW} = \frac{\Pr (\phi(\textbf{x})) }{\Pr(\phi'(1_i,\textbf{x}))}\\ 
 I_{AW} = \frac{\Pr(\phi'(0_i,\textbf{x}))}{\Pr (\phi(\textbf{x}))}
 \end{aligned}
 \end{equation}
 
\noindent Using Definitions 1-2, we formally define the above functions in HOL as follows:

 \begin{flushleft}
 	\texttt{\bf{Definition 6: }}
 	\vspace{1pt} \small{\texttt{$\vdash$ $\forall$ f i L.\\
 			I\_RW p i f L = 			$\dfrac{\texttt{prob p ($\phi$ f L)}} {\texttt{prob p ($\phi'$ (p\_space p) i f L)}}$ 			
 	}}
 \end{flushleft}
 
 \begin{flushleft}
 	\texttt{\bf{Definition 7: }}
 	\vspace{1pt} \small{\texttt{$\vdash$ $\forall$ f i L.\\
 			I\_AW p i f L = 			$\dfrac {\texttt{prob p ($\phi'$ (\{\}) i f L)}} {\texttt{prob p ($\phi$ f L)}}$ 			
 	}}
 \end{flushleft}

The HOL formalization of the above-mentioned importance measures is also available at~\cite{B_Imp_19}. The proof script of the above formalizations and proofs consists of about 1400 lines of HOL code that roughly took 70 man-hours of development time. To illustrate the effectiveness of our proposed approach, we conduct the formal importance measure analysis of a railway signaling system at Moroccan level crossing in the next section.

\section{Signaling System at Moroccan Level Crossing}	
\label{sec:imp_measure_LC}

\begin{figure}[!htb]
	\centering
	\includegraphics[scale=0.2]{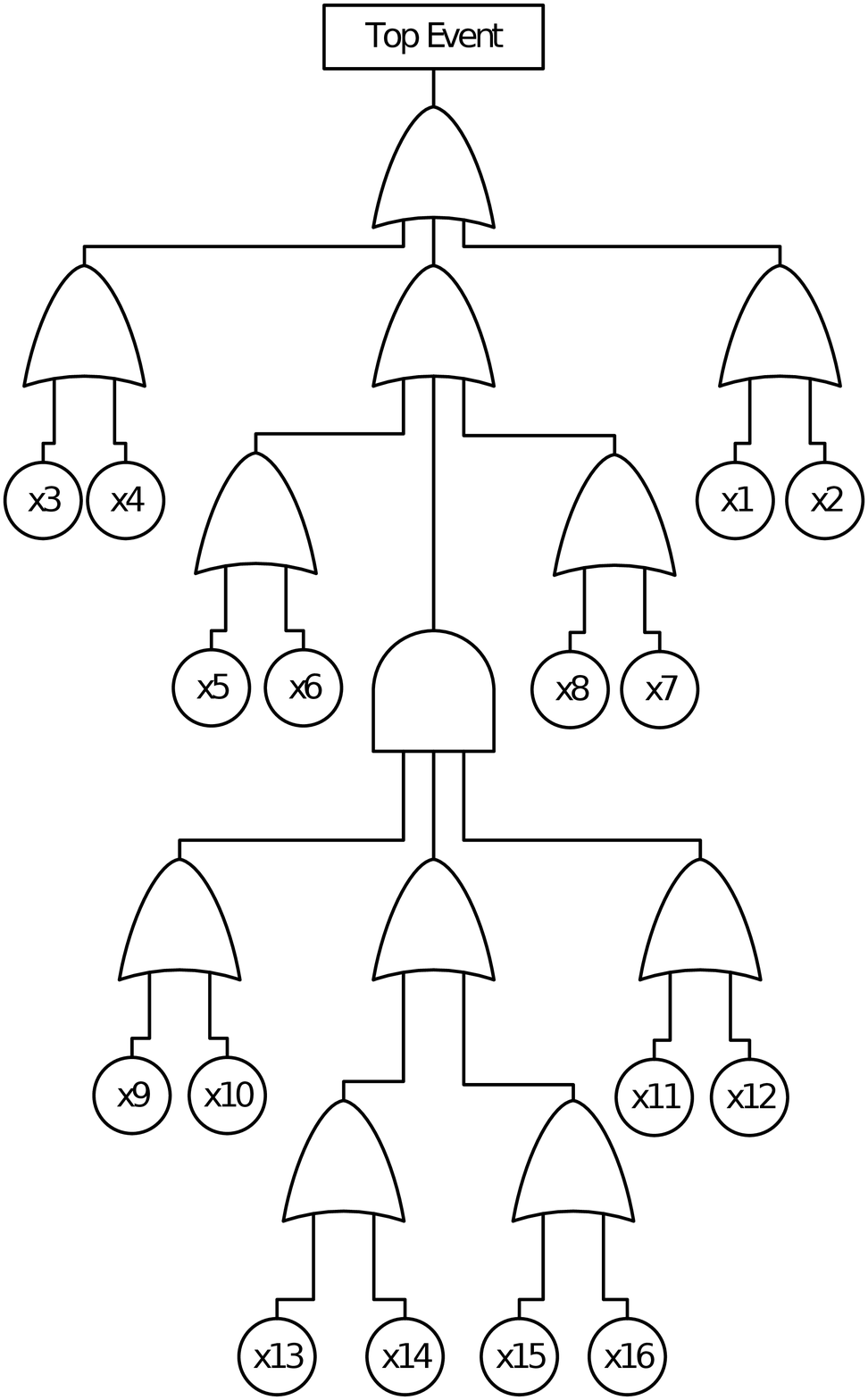}
	\caption{FT of the Signaling System at Moroccan LC}
	\label{fig:moroccon_LC}
\end{figure}

There are three main parts in the Moroccan level crossing railway signaling (LC) system~\cite{boudnnaya2014dependability}: (1) Rail part consisting of a material component (train and
rail-road), and human component (the train operator); (2) Road part containing a material component (vehicle
and road), and a human component (vehicle driver); and (3) Level crossing, which is further composed of three main components:
(i) Power and  communication  network
between the components of the railway signaling system; (ii) Control component consisting of Programmable Logic Controller and its program; 
(iii) Operative component representing sensors, such as road lights, the alarms and the barriers. Table~\ref{table:moroccan} describes the basic failure events along with the corresponding failure rates ($\lambda$) associated with the components of Moroccan signaling system~\cite{boudnnaya2014dependability}. The FT diagram of the signaling system at Moroccan LC is depicted in Figure~\ref{fig:moroccon_LC}~\cite{boudnnaya2014dependability}. 

\begin{table}[!htb]
	\centering
	\caption{Events for Signaling System at Moroccan LC}
	\begin{tabular}{|l|l|l|}
		\hline
		Symbol & Basic Events & $\lambda$ ($h^{-1}$)\\ 
		\hline
		\hline
		x1 & Vehicle Failure & $18 * 10^{-3}$\\
		x2 \& x4 & Human Factor & $1.347 * 10 ^{-4}$\\
		x3 & Rail Failure & $2.85 * 10^{-6}$\\
		x5 & Program Error & $5 * 10^{-8}$\\
		x6 & Programmable Logic Controller Failure & $4 * 10^{-6}$\\
		x7 & Network Communication Failure & $5 * 10^{-6}$\\
		x8 & Power Network Failure & $5 * 10^{-6}$\\
		x9 \& x10 & Alarm Failure & $4 * 10^{-4}$\\
		x11 \& x12 & Light Failure & $4 * 10^{-4}$\\
		x13 \& x14 & Motor Failure & $3 * 10^{-6}$\\
		x15 \& x16 & Transmission System Failure & $5 * 10^{-5}$\\
		\hline
	\end{tabular} \label{table:moroccan}
\end{table}

\subsection{Formal Model and Failure Analysis}

Using the FT gates, described in Section~\ref{subsec:formal_FTs}, we can formally model the FT diagram of the Moroccan signaling system in HOL as follows:

\vspace{2mm}
\begin{flushleft}
	\texttt{\bf{Definition 8: }}
	\vspace{1pt} \small{\texttt{$\vdash$ Signal\_FT p x1 x2 $\cdots$ x16  t = \\
			FTree p (OR 	[OR ([$\omega$ p x3 t;$\omega$ p x4 t])\\
			\ \ \ \ \ \ \ \  \ \ \ \ \	  OR ([$\omega$ p x5 t;$\omega$ p x6 t]));\\
			\ \ \ \ \ \ \ \  \ \ \ \ \		    AND [OR ([$\omega$ p x9 t;$\omega$ p x10 t]));\\
			\ \ \ \ \ \ \ \  \ \ \ \ \	\ \ \ \	\        OR ([$\omega$ p x13 t;$\omega$ p x14 t]));\\
			\ \ \ \ \ \ \ \  \ \ \ \ \	\ \ \ \	\  		OR ([$\omega$ p x15 t;$\omega$ p x16 t]));\\
			\ \ \ \ \ \ \ \  \ \ \ \ \	\ \ \ \  		OR ([$\omega$ p x11 t;$\omega$ p x12 t]))];\\
			\ \ \ \ \ \ \ \  \ \ \ \ \		    OR ([$\omega$ p x7 t;$\omega$ p x8 t]));\\
			\ \ \ \ \ \ \ \  \ \ \ \ \		    OR ([$\omega$ p x1 t;$\omega$ p x2 t]))])
	}}
\end{flushleft}

\noindent where $\omega \ p \ x \ t$ represent various failure events, such as an alarm, associated with the various component of the Moroccan signaling system. It is defined in HOL as \texttt{PREIMAGE x \{y $|$ y $\le$ t\} $\cap$ p\_space p}~\cite{WAhmad_SETTA}. 

Now, we obtain the minimal cut sets (MCS) of the above FT model by utilizing some set properties, like distribution of intersection over union and idempotent law of intersection~\cite{international2006iec}.
 
\begin{equation}
\begin{aligned}
&C_1=\{x3,x4,x5,x6\}, C_2 =\{x9,x13,x15,x11\}, \cdots,\\ 
&C_{17} = \{x10,x14,x16,x12\}, C_{18} = \{x7,x8,x1,x2\}
\end{aligned}
\end{equation}

We can also formally verify the equivalence of the obtained signaling system MCS with the orignal FT model as follows:
\vspace{2mm}
\begin{flushleft}
	\texttt{\bf{Lemma 2: }}
	\vspace{1pt} \small{\texttt{$\vdash$ Signal\_FT p x1 x2 $\cdots$ x16  t =\\
	($\lambda$b.\\
	\ \ \ FTree p((OR of\\
	\ \ \  \ ($\lambda$a. AND (gate\_list a))) b))\\
	\ \ \ \ \ \ [$\omega$L p [x3;x4;x5;x6] t;\\
	\ \ \ \ \ \ \ $\omega$L p [x9;x13;x15;x11] t;$\cdots$;\\
	\ \ \ \ \ \ \ $\omega$L p  [x10;x14;x16;x12] t;\\
	\ \ \ \ \ \ \ $\omega$L p [x7;x8;x1;x2] t]
	}}
\end{flushleft}

\noindent where the function \texttt{$\omega$L p L t} returns the list of events by mapping the function \texttt{$\omega$ p x t}, described in Definition 8, on each element of the given list of random variables.

By using the above lemma and  Definition 8, the failure probability of the Moroccan signaling system can be formally verified in HOL as follows:
\begin{flushleft}
	\texttt{\bf{Theorem 6: }}
	\label{thm:signal_FT}
	\vspace{1pt} \small{\texttt{$\vdash$ $\forall$  p x1 x2 $\cdots$ x16 c1 c2 $\cdots$ c16 t.\\  \
			[A1]: 0 $\le$ t $\wedge$\\ \
			[A2]: FT\_conds p [x1;x2;$\cdots$;x16] t\\ \
			[A3]: exp\_dist\_list p [x1;x2;$\cdots$;x16] \\ \ \
			\ \  \ \ \ \  [c1;c2;$\cdots$;c16] $\Rightarrow$ \\
			\     (prob p (Signal\_FT p x1 x2 $\cdots$ x16  t) =\\ \vspace{1pt}
			1 - $e^{-(\lambda_{c3}t)}$ * $e^{-(\lambda_{c4}t)}$ * $e^{-(\lambda_{c5}t)}$ * $e^{-(\lambda_{c6}t)}$ *\\ \ \ \ \
			$e^{-(\lambda_{c7}t)}$ * $e^{-(\lambda_{c8}t)}$ * $e^{-(\lambda_{c1}t)}$ * $e^{-(\lambda_{c2}t)}$ *\\
			(1 - (1 - $e^{-(\lambda_{c9}t)}$ * $e^{-(\lambda_{c10}t)}$ *\\
			\ (1 - $e^{-(\lambda_{c13}t)}$ * $e^{-(\lambda_{c14}t)}$) *\\
			\ (1 - $e^{-(\lambda_{c15}t)}$ * $e^{-(\lambda_{c16}t)}$) *\\
			\ (1 - $e^{-(\lambda_{c11}t)}$ * $e^{-(\lambda_{c12}t)}$)))
	}}
\end{flushleft}

\noindent where the function \texttt{exp\_dist\_list} takes a list of random variables and a list of failure rates and makes sure that each random variable is exponentially distributed and assigned with its corresponding failure rates~\cite{WAhmad_SETTA}, i.e., \texttt{exp\_dist\_list [x1;x2] [c1;c2] = (!t. 0 $\le$ t ==> (($\Pr (\omega \ p \ x1 \ t) = 1 - e^{-\lambda_{c1}*t}$) $\wedge$ ($\Pr (\omega \ p \ x2 \ t) = 1 - e^{-\lambda_{c2}*t}$))}. The function \texttt{FT\_conds} contains two predicates \texttt{mutual\_indep} and \texttt{in\_events}, which ensure that all events associated to \texttt{rail\_signal\_FT} are mutually independent and belong to events space $p$, respectively.  The proof of Theorem~6 is based on formally verified expressions of the AND and OR FT gates, presented in Table~\ref{FT_exp_table},  and  the PIE principle described in Section~\ref{subsec:PIE}.

To evaluate Theorem 6, we wrote an ML function \texttt{auto\_signal\_morco\_FT}~\cite{B_Imp_19} that takes failure rates and time index, given in Table~\ref{table:moroccan}, and returns the following in HOL evironment:

\begin{flushleft}
	\emph{\textbf{Under the following assumptions}} \\ 
	\small{\texttt{$\vdash$
			[A1]:  0 $\le$ 5 $\wedge$\\ \
			\ [A2]: FT\_conds p [x1;x2;$\cdots$;x16] 5\\ \
			\ [A3]: exp\_dist\_list p [x1;x2;$\cdots$;x16] \\ \ \
			\ \  \ \ \ \  [0.00000285;0.00000005;$\cdots$;0.0004] $\Rightarrow$ \\
\emph{\textbf{Failure probability of Moroccan Railway Signaling System}}\\
			\     (prob p (Signal\_FT p x1 x2 $\cdots$ x16  5) =\\ \vspace{1pt}  \ \    
			0.0003494028541) }}
\end{flushleft}

\begin{figure}[!htb]
	\centering
	\includegraphics[scale=0.3]{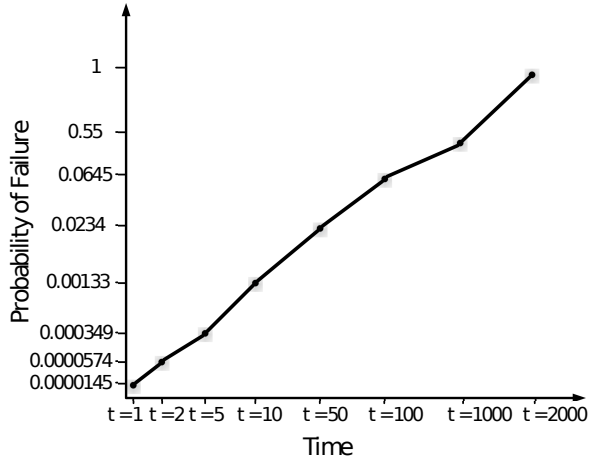}
	\caption{Plot for Probability of Failure of Signaling System at Moroccan Level Crossing}
	\label{fig:plot_LC}
\end{figure}

We can also plot these values to get a better understanding of the dependability of the Moroccan signaling system as given in Figure~\ref{fig:plot_LC}. It can be observed from the plot that initially the probability of failure is very low but as the time passes, in hours, the failure probability gradually increases and at 2,000 hours the failure becomes absolutely certain, i.e., with a probability 1.

\subsection{Formal Importance Measure Analysis}
\label{subsec:rel_impor}

As described in Section~\ref{sec:Imp_measure}, the importance measure analysis requires the given system structure function to be coherent in nature. Therefore, we start by formally verifying the conditions of a coherent system for the railway signaling system MCS, described in Lemma 2, in HOL as:  
\begin{flushleft}
	\texttt{\bf{Theorem 7: }}
	\vspace{1pt} \small{\texttt{$\vdash$\\
			prob\_space p  $\wedge$ \\
			coherent\_state\_vec (λa. a = \{\}) (FLAT \\
			\ \ \ \ \ \ [$\omega$L p [x3;x4;x5;x6] t;\\
			\ \ \ \ \ \ \ $\omega$L p [x9;x13;x15;x11] t;$\cdots$;\\
			\ \ \ \ \ \ \ $\omega$L p  [x10;x14;x16;x12] t;\\
			\ \ \ \ \ \ \ $\omega$L p [x7;x8;x1;x2] t]) $\Rightarrow$\\
			(prob p\\
			\ \ ($\phi$ ($\lambda$b.\\
			\ \ \ FTree p((OR of\\
			\ \ \  \ ($\lambda$a. AND (gate\_list a))) b))\\
			\ \ \ \ \ \ [$\omega$L p [x3;x4;x5;x6] t;\\
			\ \ \ \ \ \ \ $\omega$L p [x9;x13;x15;x11] t;$\cdots$;\\
			\ \ \ \ \ \ \ $\omega$L p  [x10;x14;x16;x12] t;\\
			\ \ \ \ \ \ \ $\omega$L p [x7;x8;x1;x2] t]) = 0)
	}}
\end{flushleft}

\begin{flushleft}
	\texttt{\bf{Theorem 8: }}
	\vspace{1pt} \small{\texttt{$\vdash$ \\
			prob\_space p  $\wedge$ \\
			coherent\_state\_vec (λa. a = p\_space p) (FLAT \\
			\ \ \ \ \ \ [$\omega$L p [x3;x4;x5;x6] t;\\
			\ \ \ \ \ \ \ $\omega$L p [x9;x13;x15;x11] t;$\cdots$;\\
			\ \ \ \ \ \ \ $\omega$L p  [x10;x14;x16;x12] t;\\
			\ \ \ \ \ \ \ $\omega$L p [x7;x8;x1;x2] t]) $\Rightarrow$\\
			(prob p\\
			\ \ ($\phi$ ($\lambda$b.\\
			\ \ \ FTree p((OR of\\
			\ \ \  \ ($\lambda$a. AND (gate\_list a))) b))\\
			\ \ \ \ \ \ [$\omega$L p [x3;x4;x5;x6] t;\\
			\ \ \ \ \ \ \ $\omega$L p [x9;x13;x15;x11] t;$\cdots$;\\
			\ \ \ \ \ \ \ $\omega$L p  [x10;x14;x16;x12] t;\\
			\ \ \ \ \ \ \ $\omega$L p [x7;x8;x1;x2] t]) = 1)
	}}
\end{flushleft}
\begin{flushleft}
	\texttt{\bf{Theorem 9: }}
	\vspace{1pt} \small{\texttt{$\vdash$\\
			prob\_space p  $\wedge$ \\
			(!t. in\_events p (FLAT\\
			\ \ \ \ \ \ [$\omega$L p [x3;x4;x5;x6] t;\\
			\ \ \ \ \ \ \ $\omega$L p [x9;x13;x15;x11] t;$\cdots$;\\
			\ \ \ \ \ \ \ $\omega$L p  [x10;x14;x16;x12] t;\\
			\ \ \ \ \ \ \ $\omega$L p [x7;x8;x1;x2] t]))  $\wedge$ \\ t1 < t2 $\Rightarrow$\\
			prob p\\
			\ ($\phi$ ($\lambda$b.\\
			\ \ \ FTree p((OR of\\
			\ \ \  \ ($\lambda$a. AND (gate\_list a))) b))\\
				\ \ \ \ \ \ [$\omega$L p [x3;x4;x5;x6] t1;\\
			\ \ \ \ \ \ \ $\omega$L p [x9;x13;x15;x11] t1;$\cdots$;\\
			\ \ \ \ \ \ \ $\omega$L p  [x10;x14;x16;x12] t1;\\
			\ \ \ \ \ \ \ $\omega$L p [x7;x8;x1;x2] t1]) $\le$\\
			prob p\\
			\ ($\phi$ ($\lambda$b.\\
			\ \ \ \ FTree p ((OR of\\
			\ \ \ \ \ \ ($\lambda$a. AND (gate\_list a))) b))\\
				\ \ \ \ \ \ [$\omega$L p [x3;x4;x5;x6] t2;\\
			\ \ \ \ \ \ \ $\omega$L p [x9;x13;x15;x11] t2;$\cdots$;\\
			\ \ \ \ \ \ \ $\omega$L p  [x10;x14;x16;x12] t2;\\
			\ \ \ \ \ \ \ $\omega$L p [x7;x8;x1;x2] t2])
	}}
\end{flushleft}

\noindent It can be seen that Theorems 7-8 are formally verified based on a very straight-forward utilization of Theorems 1-2, described in Section~\ref{subsec:coherent_sys}, on a given list of railway signaling MCS. Similary, Theorem~9 is formally verified by utilizing Theorem~3 by discharging the assumption \texttt{mem\_subset\_vec} based on the fact that by increasing the time-of-failures, i.e., \texttt{t1 $\le$ t2}, the corresponding failure probabilities also monotonically increase. 

\subsection{Formal Birnbaum Importance Measure Analysis}
After formally satisfying the conditions for coherent system on the railway signaling system failure model, we can now determine the Birnbaum importance measure of any component of the railway signaling system. For illustration purposes, we describe the formal importance measure analysis of an alarm failure ($x9$) in the railway signaling system by utilizing Definition 4 and the FT model, described in Definition 8, in HOL as:

\begin{flushleft}
	\texttt{\bf{Definition 9: }}
	\vspace{1pt} \small{\texttt{$\vdash$
			$I_\beta^9$ p x1 x2 x3 $\cdots$ x16 t =\\
			$I_\beta$ p 0\\
			\ \ ($\lambda$b. FTree p (OR 	[OR ([$\omega$ p x3 t;$\omega$ p x4 t])\\
			\ \ \ \ \ \ \ \  \ \ \ \ \	  OR ([$\omega$ p x5 t;$\omega$ p x6 t]);\\
			\ \ \ \ \ \ \ \  \ \ \ \ \		    AND [OR b ;\\
			\ \ \ \ \ \ \ \  \ \ \ \ \	\ \ \ \	\        OR ([$\omega$ p x13 t;$\omega$ p x14 t]);\\
			\ \ \ \ \ \ \ \  \ \ \ \ \	\ \ \ \	\  		OR ([$\omega$ p x15 t;$\omega$ p x16 t]);\\
			\ \ \ \ \ \ \ \  \ \ \ \ \	\ \ \ \  		OR ([$\omega$ p x11 t;$\omega$ p x12 t])];\\
			\ \ \ \ \ \ \ \  \ \ \ \ \		    OR ([$\omega$ p x7 t;$\omega$ p x8 t]);\\
			\ \ \ \ \ \ \ \  \ \ \ \ \		    OR ([$\omega$ p x1 t;$\omega$ p x2 t])]) \\ \ \ \  ([$\omega$ p x9 t;$\omega$ p x10 t]))
	}}
\end{flushleft}

The above model can also be used to quantitatively analyze the Birnbaum importance of alarm failure by associating the exponential distribution to each component of the railway signaling system as:

\vspace{3mm}
\begin{flushleft}
	\texttt{\bf{Theorem 10:}}
	\vspace{1pt} \small{\texttt{$\vdash$ $\forall$  p x1 x2 $\cdots$ x16 c1 c2 $\cdots$ c16 t.\\  \
			[A1]: 0 $\le$ t $\wedge$\\ \
			[A2]: prob\_space p $\wedge$\\ \
			[A3]: mutual\_indep p
			($\omega$L p
			\\ \ \ \ \ \ \ [x1; x2; $\cdots$; x16] t) $\wedge$\\ \
			[A4]: in\_events p 
			($\omega$L p
			\\ \ \ \ \ \ \ [x1; x2; $\cdots$; x16] t) $\wedge$\\ \
			[A5]: exp\_dist\_list p [x1;x2; $\cdots$ ;x16] \\ \ \
			\ \  \ \ \ \  [c1;c2;$\cdots$;c16] $\Rightarrow$ \\
			\ ($I_\beta^9$ p x1 x2 x3 $\cdots$ x16 t =
			\\ \ \ \ $e^{-\lambda_{c3}*t}$ * $e^{-\lambda_{c4}*t}$ * $e^{-\lambda_{c5}*t}$ *  $e^{-\lambda_{c6}*t}$ *
			\\ \ \ \   $e^{-\lambda_{c7}*t}$ * 
			$e^{-\lambda_{c8}*t}$ * $e^{-\lambda_{c2}*t}$ * $e^{-\lambda_{c1}*t}$ * $e^{-\lambda_{c10}*t}$ * 
			\\ \ \ \ \  (1 - $e^{-\lambda_{c13}*t}$ * $e^{-\lambda_{c14}*t}$) *
			\\  \ \ \ \ (1 - $e^{-\lambda_{c15}*t}$ * $e^{-\lambda_{c16}*t}$) *
			\\   \ \ \ \ (1 - $e^{-\lambda_{c11}*t}$ * $e^{-\lambda_{c12}*t}$))
	}}
\end{flushleft}
  
\noindent The assumptions of the above theorem is quite similar to the ones used in Theorem~6. It can be observed from the conclusion of Theorem~9 that the importance of alarm failure component ($x9$) is calculated from the failure probabilities of other components in the FT model.

Similarly, we can also determine the Fussell-Vesely importance measure for the alarm component by using Definition 5 in HOL as follows:

\begin{flushleft}
	\texttt{\bf{Theorem 11:}}
	\vspace{1pt} \small{\texttt{$\vdash$ $\forall$  p x1 x2 $\cdots$ x16 c1 c2 $\cdots$ c16 t.\\  \
			[A1]: 0 $\le$ t $\wedge$\\ \
			[A2]: prob\_space p $\wedge$\\ \
			[A3]: mutual\_indep p
			($\omega$L p
			\\ \ \ \ \ \ \ [x1; x2; $\cdots$; x16] t) $\wedge$\\ \
			[A4]: in\_events p 
			($\omega$L p
			\\ \ \ \ \ \ \ [x1; x2; $\cdots$; x16] t) $\wedge$\\ \
			[A5]: exp\_dist\_list p [x1;x2; $\cdots$ ;x16] \\ \ \
			\ \  \ \ \ \  [c1;c2;$\cdots$;c16] $\Rightarrow$ \\
			\ (I\_FV\_9 p x1 x2 x3 $\cdots$ x16 t =\\
			(1 -\\
			\ \ $e^{(-(\lambda_{c3} * t))}$ * $e^{(-(\lambda_{c4}* t))}$ * $e^{(-(\lambda_{c5} * t))}$ *\\
				$\cdots$\\
			(1 - $e^{(-(\lambda_{c9} * t))}$ * $e^{(-(\lambda_{c10} * t))}$)*\\
			(1 - $e^{(-(\lambda_{c13} * t))}$ * $e^{(-(\lambda_{c14} * t))}$)*\\
			(1 - $e^{(-(\lambda_{c15} * t))}$ * $e^{(-(\lambda_{c16} * t))}$)*\\
			(1 - $e^{(-(\lambda_{c11} * t))}$ * $e^{(-(\lambda_{c12}* t))}$) -\\
			(1 -\\
			\ \ $e^{(-(\lambda_{c3} * t))}$ * $e^{(-(\lambda_{c4}* t))}$ * $e^{(-(\lambda_{c5} * t))}$ *\\
			$\cdots$\\
			(1 - $e^{(-(\lambda_{c13} * t))}$ * $e^{(-(\lambda_{c14} * t))}$)*\\
			(1 - $e^{(-(\lambda_{c15} * t))}$ * $e^{(-(\lambda_{c16} * t))}$)*\\
			(1 - $e^{(-(\lambda_{c11} * t))}$ * $e^{(-(\lambda_{c12}* t))}$)) / \\
			(1 -\\
			\ \ $e^{(-(\lambda_{c3} * t))}$ * $e^{(-(\lambda_{c4}* t))}$ * $e^{(-(\lambda_{c5} * t))}$ *\\
			$\cdots$\\
			(1 - $e^{(-(\lambda_{c9} * t))}$ * $e^{(-(\lambda_{c10} * t))}$) *\\
			(1 - $e^{(-(\lambda_{c13} * t))}$ * $e^{(-(\lambda_{c14} * t))}$) *\\
			(1 - $e^{(-(\lambda_{c15} * t))}$ * $e^{(-(\lambda_{c16} * t))}$) *\\
			(1 - $e^{(-(\lambda_{c11} * t))}$ * $e^{(-(\lambda_{c12}* t))}$)))
	}}
\end{flushleft}

By using the above-mentioned approach, we can formally determine the Reduction Worth (RW) and Achievement Worth (AW) importance measures, given in Equation \ref{eq:AW_RW_imp}. Next, we conduct the formal relative importance measure analyses of relative importance among alarm and vehicle failure, using Theorem 4, as follows:

\begin{flushleft}
	\texttt{\bf{Theorem 12:}}
	\label{thm:meng_relative}
	\vspace{1pt} \small{\texttt{$\vdash$ $\forall$  p x1 x2 $\cdots$ x16 c1 c2 $\cdots$ c16 t.\\  \
			[A1]: 0 $\le$ t $\wedge$\\ \
			[A2]: prob\_space p $\wedge$\\ \
			[A3]: mutual\_indep p
			($\omega$L p
			\\ \ \ \ \ \ \ [x1; x2; $\cdots$; x16] t) $\wedge$\\ \
			[A4]: in\_events p 
			($\omega$L p
			\\ \ \ \ \ \ \ [x1; x2; $\cdots$; x16] t) $\wedge$\\ \
			[A5]: exp\_dist\_list p [x1;x2; $\cdots$ ;x16] \\
			 \	\ \  \ \ \ \  [c1;c2;$\cdots$;c16] $\wedge$\\ \ 
			[A6]: fail\_rate\_pos  [c1;c2;$\cdots$;c16] $\wedge$ \\ \ 
			[A7]: c9 $\le$ c1 $\Rightarrow$ \\
			\ \ \ $I_\beta^9$ p x1 x2 x3 $\cdots$ x16 t $\le$\\
			\ \ \ $I_\beta^1$ p x1 x2 x3 $\cdots$ x16 t
	}}
\end{flushleft}

\noindent where the function \texttt{fail\_rate\_pos}, in assumption (A6), ensures that the given list of failure rates must be positive. It can be implied from assumption (A7) that the Birnbaum relative importance of any two components in a system is related by their failure rates relationship. In other words, a component with higher failure rate is highly critical in a FT model (structure function) compared to a component with lower failure rate. The proof of Theorem 12 is based on Theorem 4 and some fundamental facts of probability theory. The HOL proof script of Theorems 6-12,  which can be downloaded from~\cite{B_Imp_19}, took about 1200 lines of HOL code and about 24 man-hours.

It is quite evident that our proposed HOL-based formalization approach provides the required rigor to the importance measure properties about system components compared to \cite{boudnnaya2014dependability}. Also, all the necessary conditions are accompanying the formally verified properties. Most importantly, the formal relative importance measure analysis reveals that the relative importance of any pair of components is related according to their failure rates (Theorem 12). In other words, we can accurately analyze the components' importance, due to the sound theorem proving approach,  without using the traditional methods of ranking the system components for large systems. By conducting the formal importance analysis of the railway signaling system at a Moroccan LC, we believe that our proposed approach provides a sound framework to reliability design engineers to meet the quality standards of their safety-critical systems.

\section {Conclusion}
\label{sec:concl}
In this paper, we formalized the commonly used importance measures, such as Birnbaum, Fussely-vesely, Reduction worth and Achievement worth, in HOL theorem proving. We also formalized Meng's approach of obtaining the relative importance measure among any pair of system components. For illustration purposes,  we conducted the formal importance measure analysis of a signaling system at a Moroccan level crossing consisting of traffic lights, programmable logic controllers and alarms, within the HOL theorem proving environment. We plan to extend the formalization of Fussell-vesely importance measure to obtain the relative importance of system components. Just like the Birnbaum importance measure, it has great potential to highlight the critical components without running the computationally expensive simulations.   
\bibliographystyle{IEEEtran}
\bibliography{biblio}

\begin{thebibliography}{10}
\providecommand{\url}[1]{#1}
\csname url@samestyle\endcsname
\providecommand{\newblock}{\relax}
\providecommand{\bibinfo}[2]{#2}
\providecommand{\BIBentrySTDinterwordspacing}{\spaceskip=0pt\relax}
\providecommand{\BIBentryALTinterwordstretchfactor}{4}
\providecommand{\BIBentryALTinterwordspacing}{\spaceskip=\fontdimen2\font plus
\BIBentryALTinterwordstretchfactor\fontdimen3\font minus
  \fontdimen4\font\relax}
\providecommand{\BIBforeignlanguage}[2]{{%
\expandafter\ifx\csname l@#1\endcsname\relax
\typeout{** WARNING: IEEEtran.bst: No hyphenation pattern has been}%
\typeout{** loaded for the language `#1'. Using the pattern for}%
\typeout{** the default language instead.}%
\else
\language=\csname l@#1\endcsname
\fi
#2}}
\providecommand{\BIBdecl}{\relax}
\BIBdecl

\bibitem{kuo2012importance}
W.~Kuo and X.~Zhu, \emph{{Importance Measures in Reliability, Risk, and
  Optimization: Principles and Applications}}.\hskip 1em plus 0.5em minus
  0.4em\relax John Wiley \& Sons, 2012.

\bibitem{rooney2002microsoft}
\BIBentryALTinterwordspacing
P.~Rooney, ``{Microsoft's CEO: 80-20 Rule Applies to Bugs, Not Just
  Features},'' 2019. [Online]. Available:
  \url{https://www.crn.com/news/security/18821726/microsofts-ceo-80-20-rule-applies-to-bugs-not-just-features.htm}
\BIBentrySTDinterwordspacing

\bibitem{birnbaum1968importance}
\BIBentryALTinterwordspacing
Z.~W. Birnbaum, ``{On the importance of Different Components in a
  Multicomponent System},'' University of Washington, Seatle, Washington, USA,
  Tech. Rep., 1968. [Online]. Available:
  \url{http://www.dtic.mil/dtic/tr/fulltext/u2/670563.pdf}
\BIBentrySTDinterwordspacing

\bibitem{espiritu2007component}
J.~F. Espiritu, D.~W. Coit, and U.~Prakash, ``{Component Criticality Importance
  Measures for the Power Industry},'' \emph{Electric Power Systems Research},
  vol.~77, no. 5-6, pp. 407--420, 2007.

\bibitem{Reliasoft_RI}
\BIBentryALTinterwordspacing
{ReliaSoft}, 2019. [Online]. Available:
  \url{https://www.weibull.com/hotwire/issue66/relbasics66.htm}
\BIBentrySTDinterwordspacing

\bibitem{boland1989optimal}
P.~J. Boland, F.~Proschan, and Y.~L. Tong, ``{Optimal Arrangement of Components
  via Pairwise Rearrangements},'' \emph{Naval Research Logistics}, vol.~36,
  no.~6, pp. 807--815, 1989.

\bibitem{meng2004comparing}
F.~C. Meng, ``{Comparing Birnbaum Importance Measure of System Components},''
  \emph{Probability in the Engineering and Informational Sciences}, vol.~18,
  no.~2, pp. 237--245, 2004.

\bibitem{harrison_09}
J.~Harrison, \emph{Handbook of {P}ractical {L}ogic and {A}utomated
  {R}easoning}.\hskip 1em plus 0.5em minus 0.4em\relax Cambridge {U}niversity
  {P}ress, 2009.

\bibitem{international2006iec}
\BIBentryALTinterwordspacing
IEC, ``International {E}lectrotechnical {C}ommission, 61025 {F}ault {T}ree
  {A}nalysis,'' 2006. [Online]. Available:
  \url{https://webstore.iec.ch/publication/4311}
\BIBentrySTDinterwordspacing

\bibitem{boudnnaya2014dependability}
J.~Boudnnaya, A.~Mkhida, and M.~Sallak, ``{A Dependability Analysis of a
  Moroccan Level Crossing Based on Fault Tree Analysis and Importance
  Measures},'' in \emph{{MOdeling, Optimization and SIMlation}}, 2014, pp.
  1--5,
  \url{https://recif.hds.utc.fr/wp-content/uploads/2014/11/MOSIM_2014_1.pdf}.

\bibitem{hol_tp}
\BIBentryALTinterwordspacing
{HOL Interactive Theorem Prover}, 2019. [Online]. Available:
  \url{https://hol-theorem-prover.org/}
\BIBentrySTDinterwordspacing

\bibitem{ahmad_phd_17}
W.~Ahmad, ``{Formal Dependability Analysis using Higher-order-logic Theorem
  Proving},'' Ph{D} {T}hesis, National University of Sciences and Technology,
  Islamabad, Pakistan, 2017.

\bibitem{Trivedi_02}
K.~S. Trivedi, \emph{{Probability and Statistics with Reliability, Queuing and
  Computer Science Applications}}.\hskip 1em plus 0.5em minus 0.4em\relax John
  Wiley and Sons Ltd., 2002.

\bibitem{ahmed2016formalization}
W.~Ahmed, O.~Hasan, and S.~Tahar, ``{Formalization of Reliability Block
  Diagrams in Higher-order Logic},'' \emph{Journal of Applied Logic}, vol.~18,
  pp. 19--41, 2016.

\bibitem{ahmad_railway_trac}
W.~Ahmad, O.~Hasan, and S.~Tahar, \emph{{Handbook of RAMS in Railways: Theory
  and Practice}}.\hskip 1em plus 0.5em minus 0.4em\relax Taylor and Francis,
  2018, ch. {Formal Reliability Analysis of Railway Systems using Theorem
  Proving Technique}, pp. 651--668.

\bibitem{CICM_15_WAhmed}
W.~Ahmad and O.~Hasan, ``{Towards Formal Fault Tree Analysis Using Theorem
  Proving},'' in \emph{Intelligent Computer Mathematics}, ser. LNCS.\hskip 1em
  plus 0.5em minus 0.4em\relax Springer, 2015, vol. 9150, pp. 39--54.

\bibitem{WAhmad_SETTA}
W.~Ahmad and O.~Hasan, ``{Formalization of Fault Trees in Higher-order Logic: A
  Deep Embedding Approach},'' in \emph{Dependable Software Engineering:
  Theories, Tools, and Applications}, ser. LNCS.\hskip 1em plus 0.5em minus
  0.4em\relax Springer, 2016, vol. 9984, pp. 264--279.

\bibitem{gordon1993introduction}
M.~J. Gordon and T.~F. Melham, \emph{{Introduction to HOL A Theorem Proving
  Environment for Higher-order Logic}}.\hskip 1em plus 0.5em minus 0.4em\relax
  Cambridge University Press, 1993.

\bibitem{church_40}
A.~Church, ``A {F}ormulation of the {S}imple {T}heory of {T}ypes,''
  \emph{Journal of {S}ymbolic {L}ogic}, vol.~5, pp. 56--68, 1940.

\bibitem{milner_77}
R.~Milner, ``A {T}heory of {T}ype {P}olymorphism in {P}rogramming,''
  \emph{Journal of {C}omputer and {S}ystem {S}ciences}, vol.~17, pp. 348--375,
  1977.

\bibitem{mhamdi_11}
T.~Mhamdi, O.~Hasan, and S.~Tahar, ``{O}n the {F}ormalization of the {L}ebesgue
  {I}ntegration {T}heory in {HOL},'' in \emph{Interactive {T}heorem {P}roving},
  ser. {LNCS}.\hskip 1em plus 0.5em minus 0.4em\relax Springer, 2011, vol.
  6172, pp. 387--402.

\bibitem{WAhmad_CICM14}
W.~Ahmad, O.~Hasan, S.~Tahar, and M.~S. Hamdi, ``{Towards the Formal
  Reliability Analysis of Oil and Gas Pipelines},'' in \emph{Intelligent
  Computer Mathematics}, ser. LNCS.\hskip 1em plus 0.5em minus 0.4em\relax
  Springer, 2014, vol. 8543, pp. 30--44.

\bibitem{andrews2003birnbaum}
J.~D. Andrews and S.~C. Beeson, ``{Birnbaum's Measure of Component Importance
  for Noncoherent Systems},'' \emph{IEEE Transcation on Reliability}, vol.~52,
  no.~2, pp. 213--219, 2003.

\bibitem{B_Imp_19}
\BIBentryALTinterwordspacing
W.~Ahmad, ``{On the Formalization of Importance Measures using HOL Theorem
  Proving},'' 2019. [Online]. Available:
  \url{https://github.com/ahmedwaqar/Formal-Dependability/tree/importance_measures/Importance_Measures}
\BIBentrySTDinterwordspacing

\end{thebibliography}
	

	%
	\IEEEpeerreviewmaketitle

\end{document}